\documentclass{appolb}
\usepackage{graphicx}
\usepackage{amsmath}
\usepackage{amssymb}
\usepackage{amsfonts}
\usepackage{xcolor}
\usepackage{hyperref,physics,wrapfig,lineno}

\headtitle{Conserved charge fluctuations in the chiral limit}
\headauthor{Mugdha Sarkar et al}
\begin{document}
\title{Conserved charge fluctuations in the chiral limit%
\thanks{Presented at workshop on Criticality in QCD and the Hadron Resonance Gas; 29-31 July 2020, Wroclaw, Poland.}%
}
\author{Mugdha Sarkar\thanks{Speaker}, Olaf Kaczmarek, Frithjof Karsch, Anirban Lahiri, Christian Schmidt
\address{Fakult\"at f\"ur Physik, Universit\"at Bielefeld, D-33615 Bielefeld, Germany.}
}
\maketitle
\begin{abstract}
We study the signs of criticality in conserved charge fluctuations and related observables of finite temperature QCD at vanishing chemical potential, as we approach the chiral limit of two light quarks. Our calculations have been performed on gauge ensembles generated using Highly Improved Staggered Quark (HISQ) fermion action, with pion masses ranging from $140$ MeV to $55$ MeV. 
\end{abstract}
\PACS{11.10.Wx, 11.15.Ha, 12.38.Aw, 12.38.Gc, 12.38.Mh, 24.60.Ky, 25.75.Nq}
  
\section{Introduction}
\begin{wrapfigure}{r}{0.45\textwidth}
  \begin{center}
    \includegraphics[width=5cm]{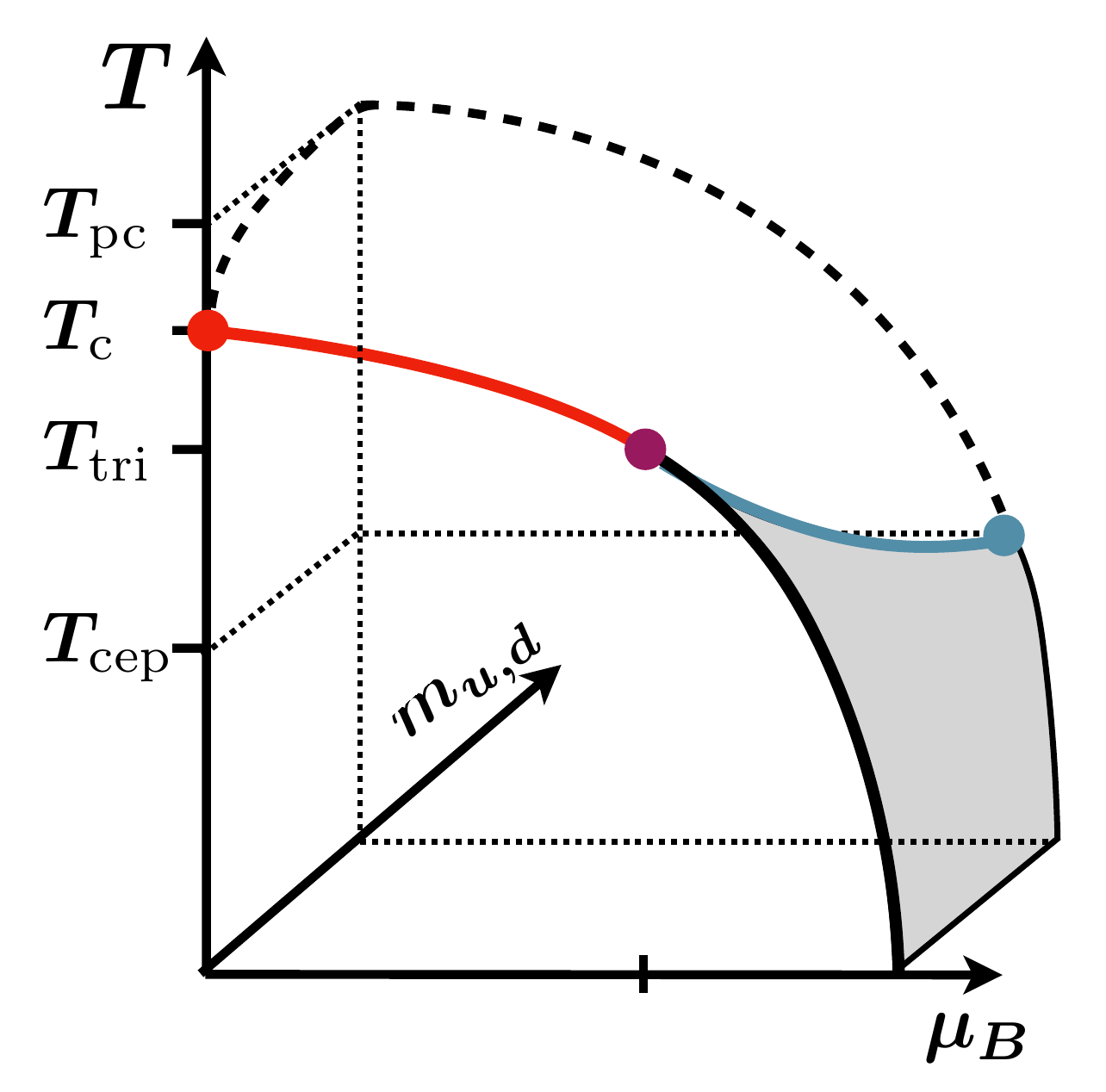}
  \end{center}
  \caption{Schematic phase diagram of QCD in the space of temperature $T$, baryon chemical potential $\mu_B$ and light quark masses $m_{u,d}$ \cite{Karsch:2019mbv}.}
  \label{Fig:phasediag}
\end{wrapfigure}

Understanding the phase diagram of QCD in the plane of temperature and various chemical potentials is one of the primary goals of lattice QCD calculations and the heavy-ion collision experiments at RHIC and LHC. The temperature variation at zero chemical potential is being explored at LHC and studied extensively using lattice techniques due to the absence of the infamous sign problem. At physical quark masses, there exists a chiral crossover at a temperature $T_{pc}$ around $157$ MeV \cite{Bazavov:2018mes, Borsanyi:2020fev}. A schematic phase diagram of QCD with an additional axis for degenerate
light (up and down) quark masses is shown in Fig. \ref{Fig:phasediag} \cite{Karsch:2019mbv}. In the figure, the dotted lines along the horizontal and vertical axes correspond to the plane of physical quark masses and the dashed line is the chiral crossover line which ends at a critical end point $T_{cep}$, which is being actively pursued in experiments. 

The chiral limit, however, is not accessible to experiments and can only be studied via theoretical techniques. The spontaneous symmetry breaking of the exact $SU(2)\times SU(2)$ symmetry in the chiral limit of two light quarks is expected to be a phase transition belonging to the universality class of $3d$ $O(4)$ spin model \cite{Pisarski:1983ms}.
There exists an alternative scenario where the chiral crossover turns into a first order transition as we move towards 
the chiral limit, through a $Z(2)$ critical point at an intermediate light quark mass value \cite{Pisarski:1983ms,Philipsen:2016hkv}. 
The order and nature of the chiral phase transition is still not clear beyond doubt. Recently, the phase transition temperature $T_c$ was found to be $132^{+3}_{-6}$ MeV in Ref.~\cite{Ding:2019prx}. {Recent works have shown evidences which imply that the $O(4)$ scenario is favored \cite{Ding:2019prx,Ding:2018auz,Clarke:2020htu,alahiriproc2020}}. In this work, we thus consider only the $O(4)$ universality class and try to understand the imprint of this criticality on thermodynamic observables, in particular on conserved charge fluctuations. The importance of these effects in the cumulants with regard to experiments have been discussed in Ref.~\cite{Friman:2011pf}. 

The plan of this proceedings is as follows. After the brief introduction, Sec. \ref{sec2} discusses the theoretical framework of studying universal critical phenomena and introduces the observables we are interested in. Our numerical setup is briefly summarized in Sec. \ref{Sec3} followed by our results in Sec. \ref{Sec4}. We conclude and present our future directions in Sec. \ref{Sec5}. 

\section{Critical behavior in the chiral limit} \label{sec2}
According to Wilson's renormalization group (RG) theory, the effective Hamiltonian of a theory in the space of all possible couplings consists of ``energy-like'' terms which respect the symmetry and ``magnetic-like'' terms which break the symmetry. In QCD, the temperature $T$ and chemical potentials $\mu_X$ for different conserved charges $X=$ baryon number $B$, electric charge $Q$, strangeness $S,\ldots$ would thus be energy-like couplings with respect to the chiral phase transition whereas the light quark mass $m_l\equiv m_{u,d}$ would be the magnetic-like coupling.

In the vicinity of a phase transition, the imprint of the criticality in thermodynamic quantities can be expressed as singular or non-analytic universal contributions. The starting point of the discussion is writing down the logarithm of the partition function i.e. the free energy density or the pressure, close to a phase transition using generalized scaling laws \cite{Engels:2011km},
\begin{equation}\label{Eq:free_energy}
 \frac{1}{T^4}f(T,\vec{\mu},m_l)= \frac{1}{VT^3}\ln Z(T,\vec\mu,m_l) = { h^{(2-\alpha)/\beta\delta}f_f(z)} + f_r(T,\vec{\mu},m_l),
\end{equation}
where $V$ is the system volume and $f_r$ denotes the regular non-critical contributions which are particular to the theory, QCD in our case. The non-analytic contribution is expressed in terms of a generalized energy-like coupling, the reduced temperature $t$ and a magnetic-like coupling $h$, written upto leading order near the critical point, as
\begin{equation}\label{Eq:th}
 t=\frac{1}{t_0}\left(\frac{T-T_c}{T_c} + \kappa_2^X\left(\frac{\mu_X}{T}\right)^2\right), \quad h=\frac{1}{h_0}\frac{m_l}{m_s}.
\end{equation}
These dimensionless couplings are defined such that the phase transition occurs at $t=h=0$.
Since we are interested in the chiral phase transition at chemical potential $\mu_X=0$, this
corresponds to temperature $T=T_c$ and light quark mass $m_l=0$. The strange quark mass $m_s$ 
is used in $h$ to get rid of multiplicative mass renormalization factors in order to obtain a well-defined scaling field in the continuum limit.
It is important to note that $m_s$ does not break the two-flavor chiral symmetry group and in principle, can also be included in the definition of $t$. The dimensionless $t_0$ and $h_0$ are non-universal constants. In the singular term in Eq. \ref{Eq:free_energy}, $f_f(z)$ is a universal scaling function with the scaling variable $z$ being a particular combination of the $t$ and $h$ couplings, $z\equiv t/h^{1/\beta\delta}$. Depending on the universality class, the critical exponents $\alpha,\beta$ and $\delta$ determine the singular behavior in the chiral limit, $h\to 0$. 

\begin{figure}[t]
\centerline{%
\includegraphics[width=5cm]{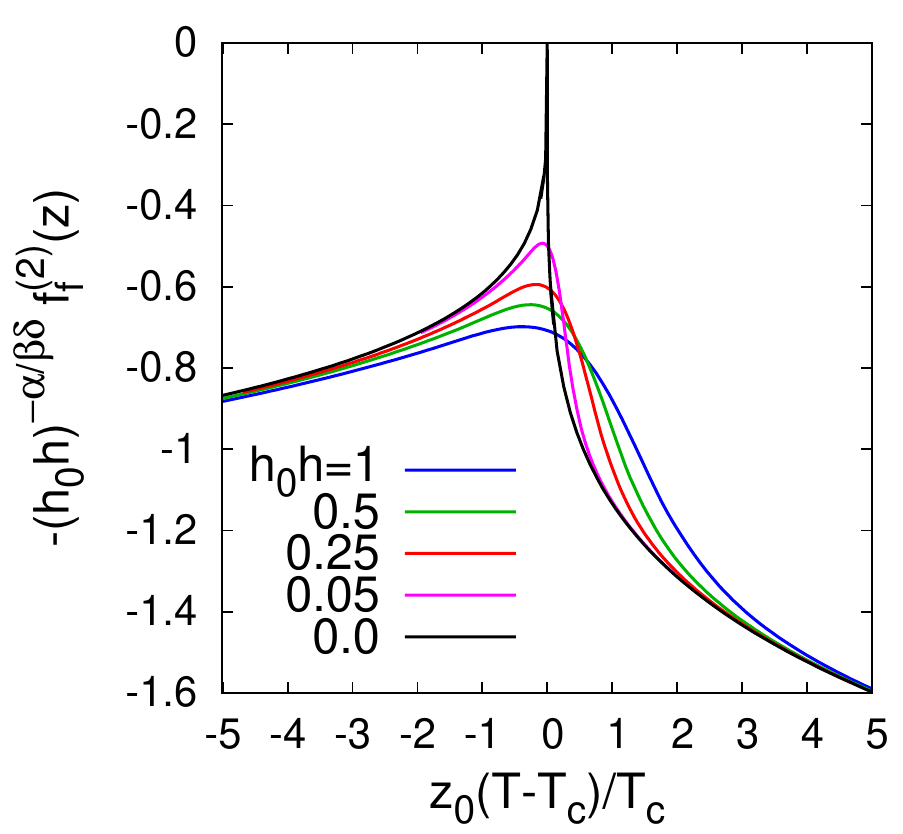}
\includegraphics[width=5cm]{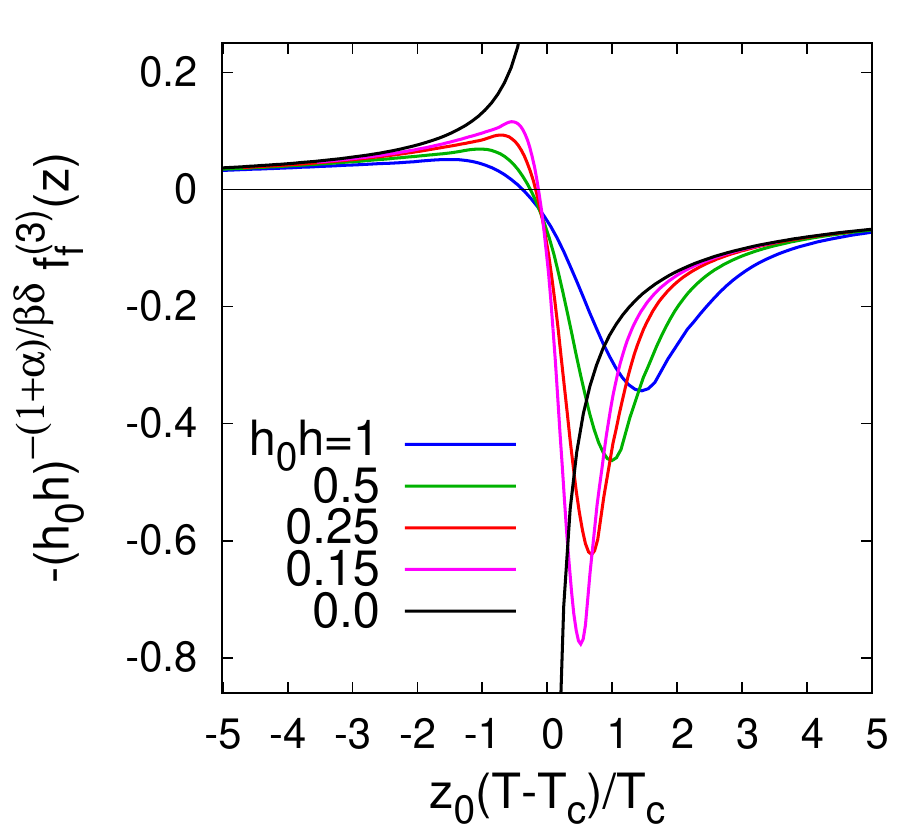}}
\caption{Scaling behavior of the scaled singular parts of fourth (left) and sixth(right) order fluctuations as a function of the reduced temperature scaled by $z_0\equiv h_0^{1/\beta\delta}/t_0$. The figure has been taken from Ref.~\cite{Friman:2011pf}}
\label{Fig:o4func}
\end{figure}

The conserved charge fluctuations are obtained as derivatives of the free energy density $f$ w.r.t. the corresponding chemical potentials and are therefore, energy-like observables. At zero chemical potential, the expressions for the singular parts in these cumulants can be obtained from Eq. \ref{Eq:free_energy} as
\begin{equation}
 {\chi_{2n}^X} = -\frac{\partial^{2n}f/T^4}{\partial(\mu_X/T)^{2n}}\Bigg|_{\mu_X=0}\sim \quad {-~(2\kappa^X_2)^n ~h^{(2-\alpha-n)/\beta\delta}f_f^{(n)}(z)},
\end{equation}
where $f_f^{(n)}(z)$ is the $n$th derivative w.r.t the scaling variable $z$.
Note that the odd order cumulants are zero. From Eq. \ref{Eq:th}, it is easy to see that a single temperature derivative yields a singular part same as the one obtained after a double chemical potential derivative. Therefore, if the regular contributions are small enough, the second and fourth order conserved charge fluctuations would behave like energy density and specific heat, respectively.
As one expects the chiral phase transition to belong to the $O(4)$ universality class, one can plug in the $O(4)$ critical exponents in the above-mentioned cumulants. Since $\alpha$ is negative\footnote{The fourth order cumulant would diverge for the $Z(2)$ universality class as $\alpha$ is positive.} one finds that the second and fourth order fluctuations remain finite in the chiral limit whereas the sixth and higher order fluctuations diverge. The scaling behavior of the universal non-analytic contributions of $\chi_4$ and $\chi_6$ obtained from $3d$ $O(4)$ spin model calculations, are shown in Fig. \ref{Fig:o4func}, which have been taken from Ref.~\cite{Friman:2011pf}. The plots show the variation of the singular parts w.r.t. the reduced temperature for different $h$. It can be clearly seen in the left plot that the singular part of $\chi_4$ does not diverge in the chiral limit $h\to 0$ but instead becomes zero at $T_c$ with a characteristic spike. On the other hand, the sixth order cumulant $\chi_6$ has a positive and negative peak which diverge in the chiral limit. We will confront these expectations with our numerical findings.

It is also interesting to look at the ``mixed'' observables which are derivatives of the free energy with respect to both the energy-like and magnetic-like couplings. In contrast to energy-like observables, mixed quantities are divergent already at second order. For example, consider the second order conserved charge fluctuation, 
\begin{equation}\label{Eq:chi2X}
 \chi_2^X = -A\kappa^X_2 H^{(1-\alpha)/\beta\delta}f_f^\prime(z) + \chi^X_{2,\mathrm{reg}},
\end{equation}
where we move the factor of non-universal constant $h_0$ into $A$ by defining $H\equiv m_l/m_s = h_0 h$ and $\chi^X_{2,\mathrm{reg}}$ denotes the regular terms. Taking a $H$-derivative, one obtains the second order mixed susceptibility
\begin{equation}\label{Eq:mixed}
 \frac{\partial \chi_2^X}{\partial H} = A\kappa^X_2 H^{(\beta-1)/\beta\delta}f_G^\prime(z) + \frac{\partial \chi^X_{2,\mathrm{reg}}}{\partial H}
\end{equation}
where $f_G$ is a universal scaling function related to $f_f$ as $f_G(z)=-(1+ 1/\delta)f_f(z)+(z/\beta\delta)f_f^\prime(z)$. The observable 
has a moderate divergence, i.e., $H^{(\beta-1)/\beta\delta} = H^{-0.34}$ in the chiral limit.

\section{Numerical setup} \label{Sec3}
Our calculations have been done on gauge ensembles generated by the HotQCD collaboration with the Highly Improved Staggered Quark (HISQ) fermion discretization and tree-level Symanzik-improved gauge action. Part of these ensembles have been recently used in the determination of the chiral phase transition temperature $T_c$ \cite{Ding:2019prx} and in studying the sensitivity of the Polyakov loop to the chiral phase transition \cite{Clarke:2020htu}. Keeping the strange quark mass $m_s$ fixed at its physical value, the gauge configurations have been generated with light quark masses $m_l=m_s/27, m_s/40, m_s/80$ and $m_s/160$, corresponding to pion masses $140$ MeV, 110 MeV, 80 MeV and 55 MeV respectively. The measurements have been done in the largest available volumes at one value of lattice spacing, $a=1/8T$, i.e., at fixed lattice temporal extent $N_\tau=8$. 

\section{Results} \label{Sec4}
\begin{figure}[t]
\centerline{%
\includegraphics[width=6cm]{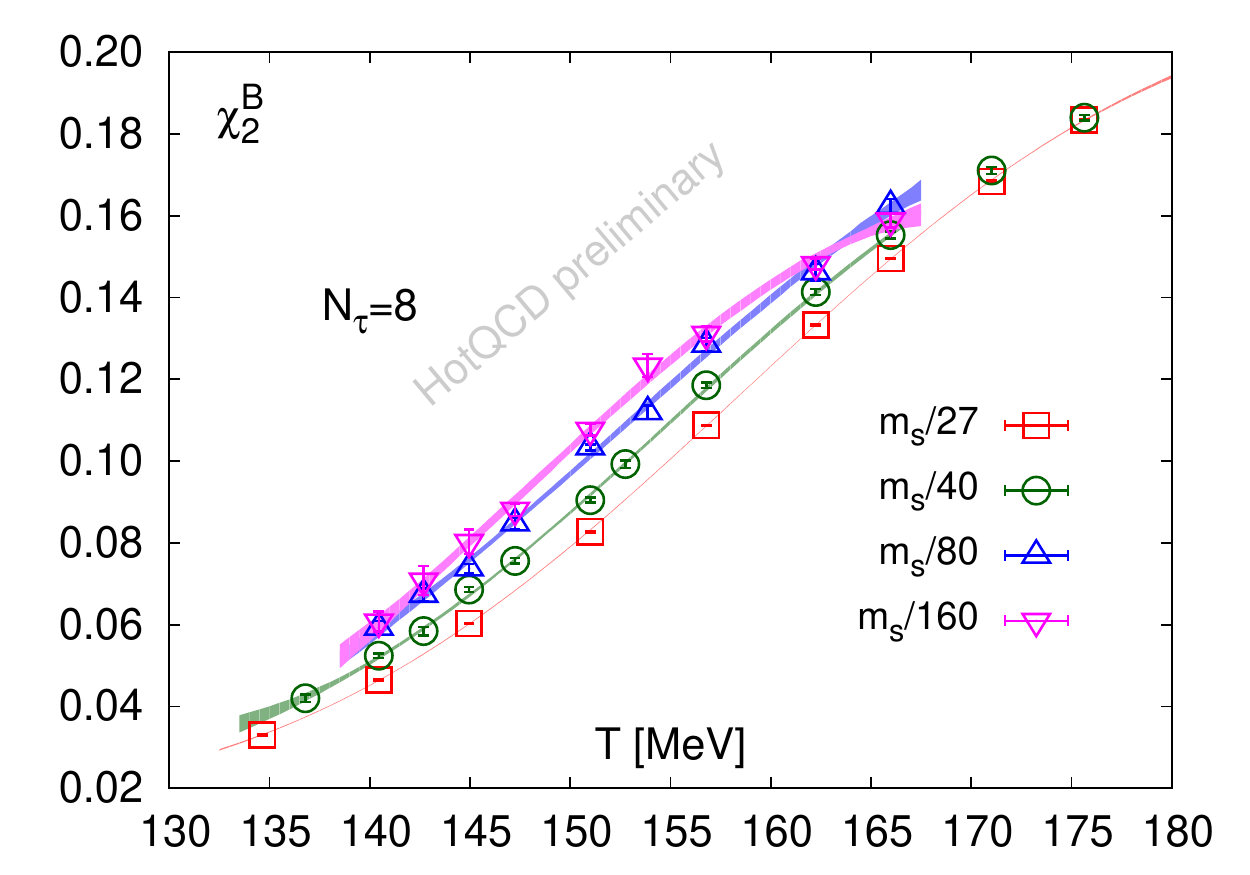}
\includegraphics[width=6cm]{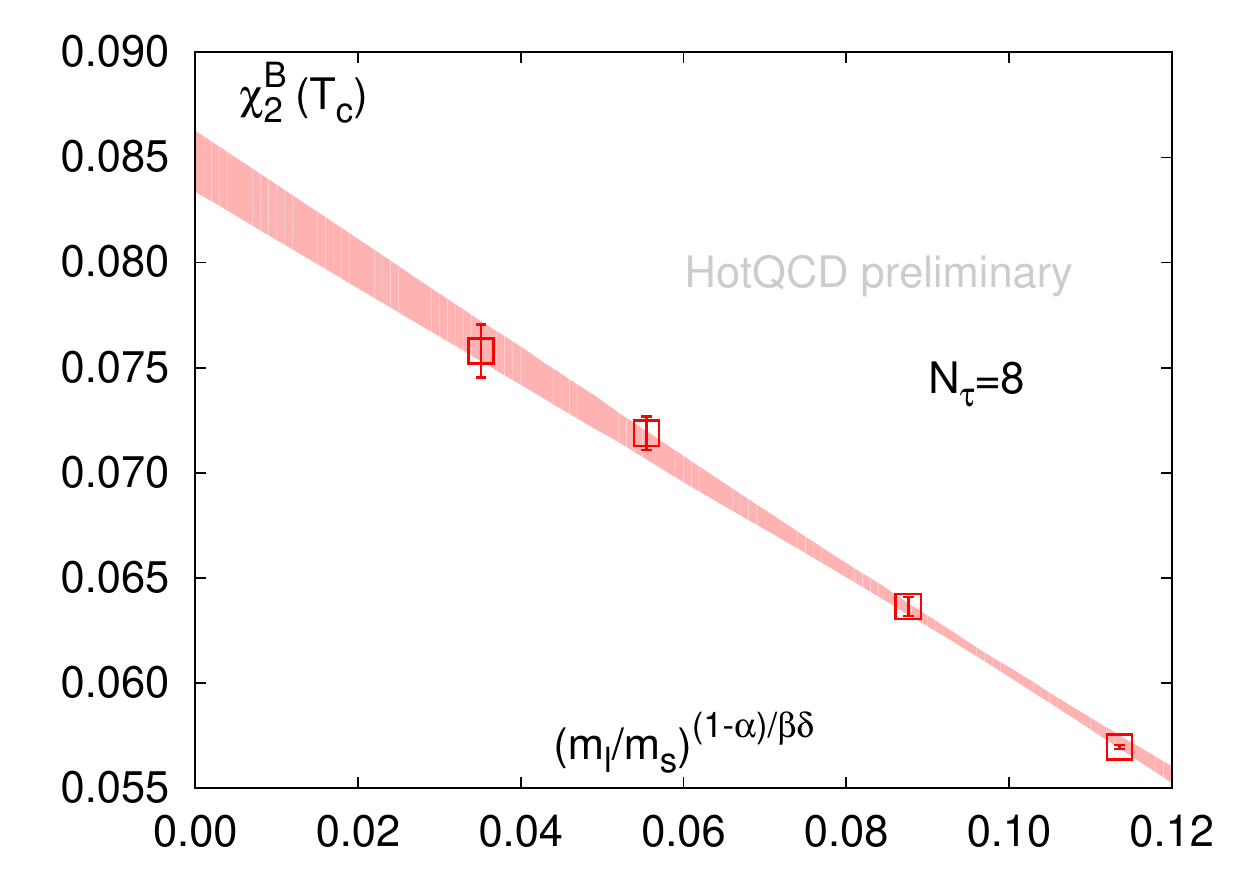}}
\caption{(\textit{Left}) Second order net baryon-number fluctuations as a function of temperature for different $H$ values. (\textit{Right}) Linear fit of $\chi_2^B(T_c^{N_\tau=8})$ as a function of $H^{(1-\alpha)/\beta\delta}$ using $O(2)$ critical exponents.}
\label{Fig:chi2B}
\end{figure}
{At the outset, we mention that since our work has been done with staggered quarks at finite lattice spacing, the relevant universality class would be that of $3d$ $O(2)$ spin models. Hence, we use the $O(2)$ critical exponents which are quite close to $O(4)$ and the qualitative conclusions should remain the same.}

\begin{figure}[t]
\centerline{%
\includegraphics[width=6cm]{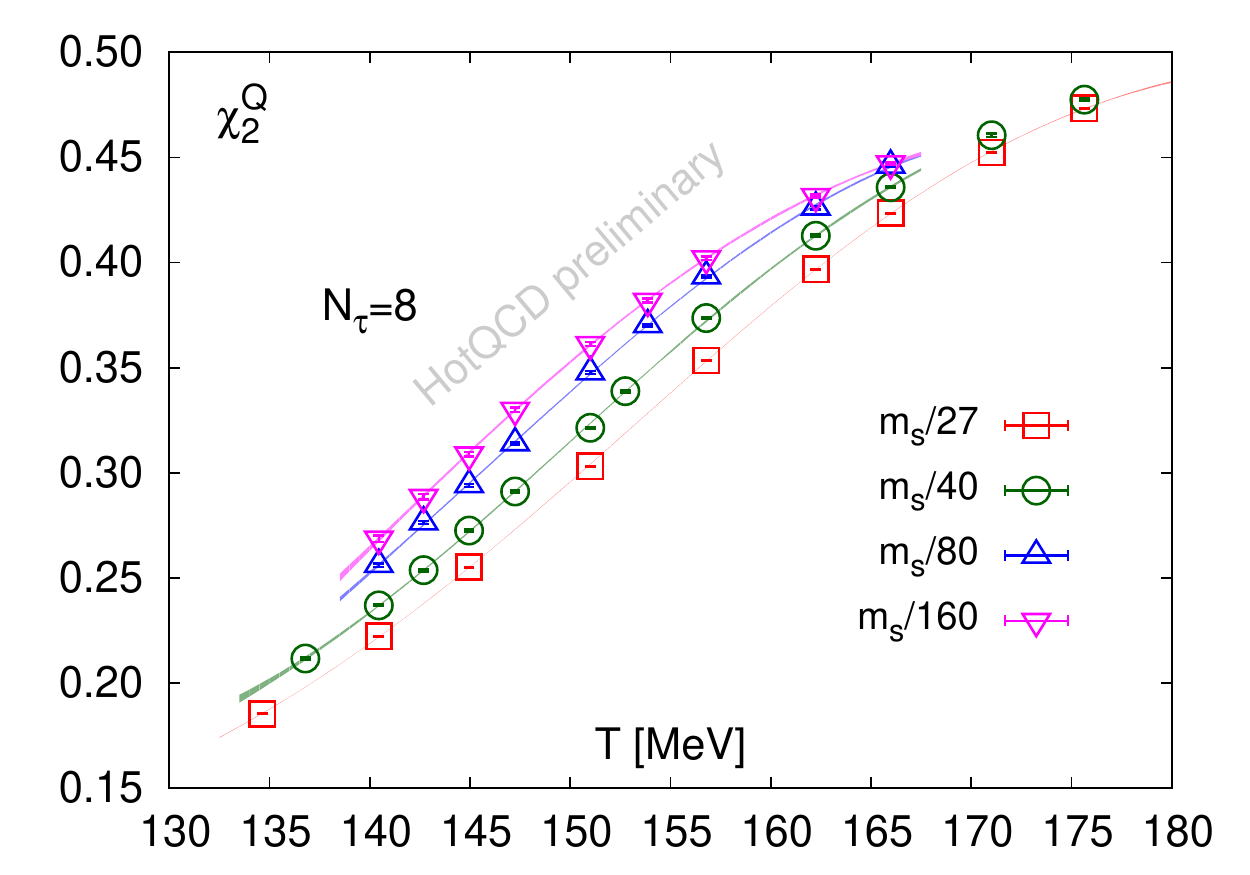}
\includegraphics[width=6cm]{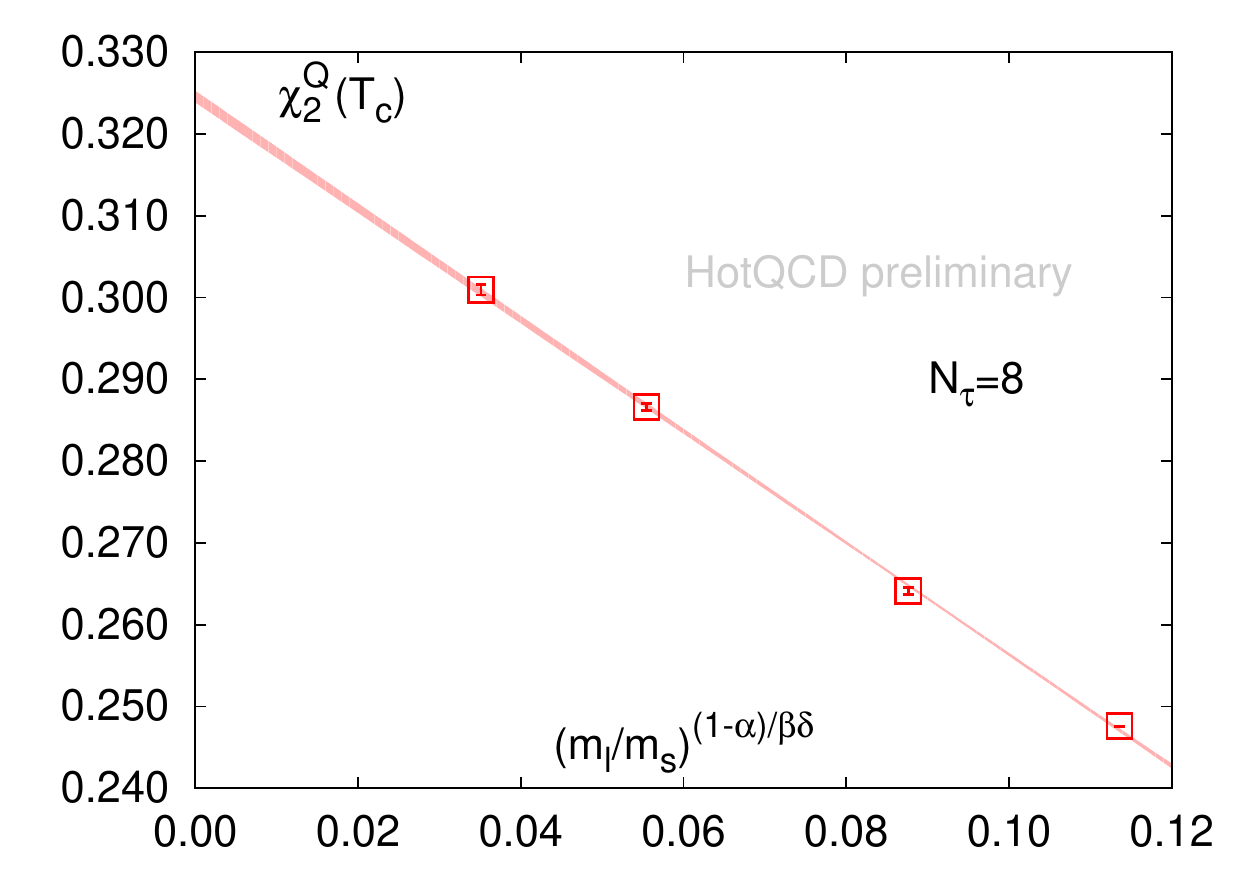}}
\caption{Same as Fig. \ref{Fig:chi2B} but for electric charge fluctuations}
\label{Fig:chi2Q}
\end{figure}
\begin{figure}[t]
\centerline{%
\includegraphics[width=6cm]{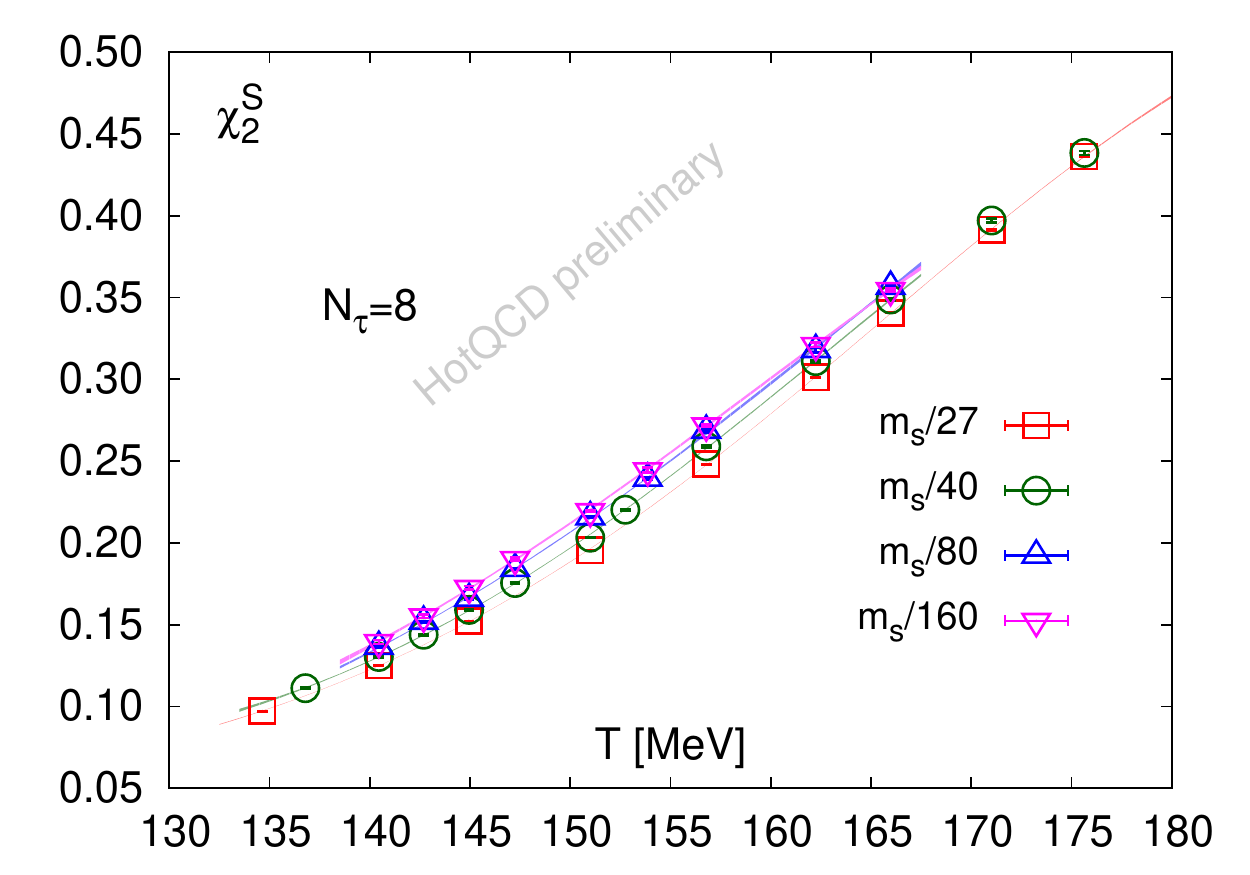}
\includegraphics[width=6cm]{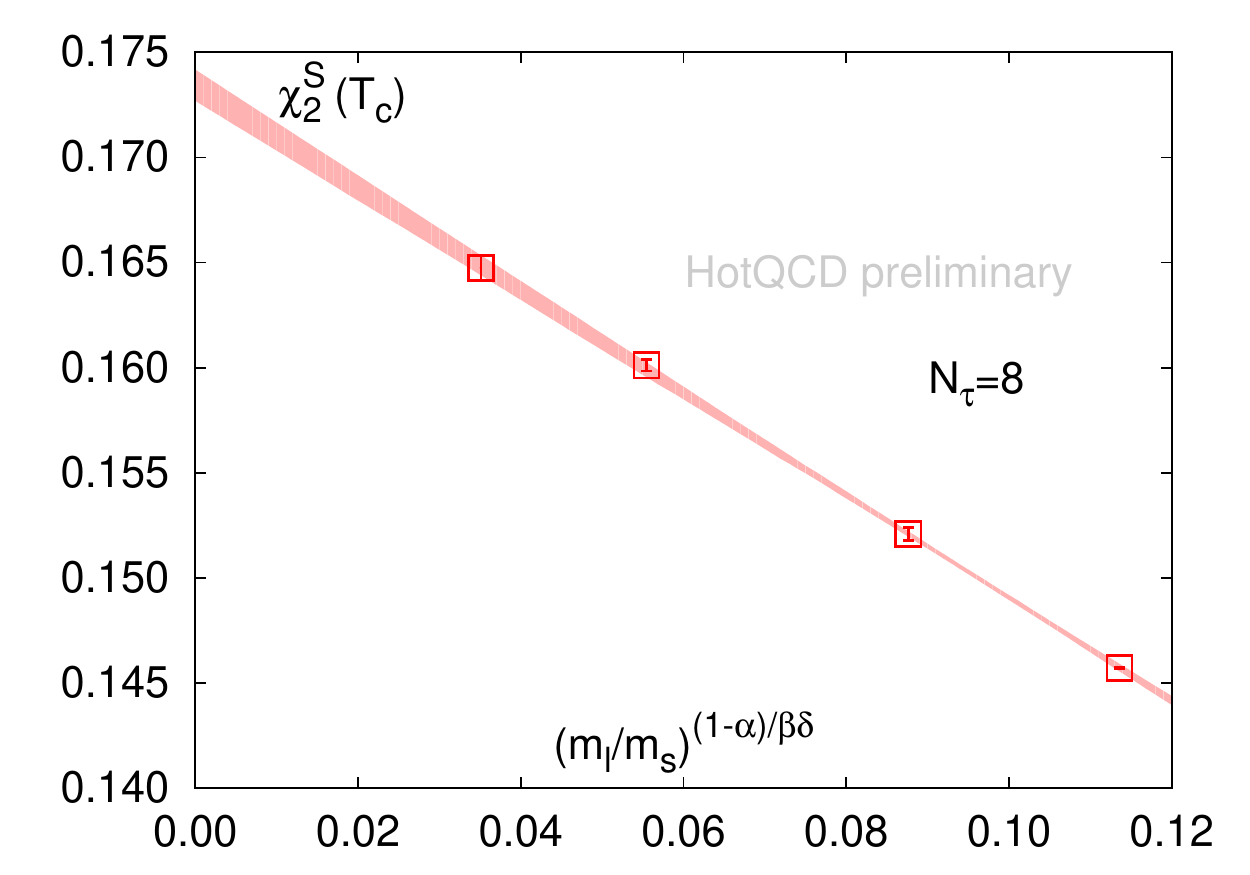}}
\caption{Same as Fig. \ref{Fig:chi2B} but for strangeness fluctuations}
\label{Fig:chi2S}
\end{figure}
We start with the results of second order conserved charge fluctuations, shown in the left plots of Figs. \ref{Fig:chi2B}, \ref{Fig:chi2Q} and \ref{Fig:chi2S}. In the scaling regime, the second order chemical potential derivative would behave like a single derivative w.r.t.~temperature, i.e., like an energy density, if the singular part has a dominant contribution. It is possible to estimate the singular contribution in these quantities by extrapolating the value of the function in the chiral limit at a given temperature \cite{Engels:2011km}. Setting $T=T_c$ at any given light quark mass in Eq.~\ref{Eq:chi2X}, we have 
\begin{equation}\label{Eq:chi2atTc}
 \chi_2^X(t=0,H) = -A\kappa^X_2{H^{(1-\alpha)/\beta\delta}}{f_f^{(1)}(0)} + \text{const. reg. term} + \mathcal{O}(H^2).
\end{equation}
We see that the leading mass dependence is given by the critical exponents in the singular term, followed\footnote{The free energy density is even in $H$.} by $H^2$. For $O(N)$ and $Z(2)$ universality classes, the combination $(1-\alpha)/\beta\delta$ is positive and $<1$, thus, the singular term vanishes in the chiral limit. Using the $T_c$ value extracted in \cite{Ding:2019prx} for $N_\tau=8$, we can plot $\chi_2^X(T_c,H)$ versus $H^{(1-\alpha)/\beta\delta}$, as shown in the right plots of Figs. \ref{Fig:chi2B}, \ref{Fig:chi2Q}, \ref{Fig:chi2S}. For small enough $H$, one would expect a linear dependence, which is what we find already from physical light quark masses within errors. From the intercept of the linear fit which denotes the constant regular term in Eq. \ref{Eq:chi2atTc}, one can obtain the singular contribution at physical quark mass as $\chi_2^X (T_c,H=0) - \chi_2^X(T_c,H=1/27)$, upto leading order. We have listed the singular contributions for physical quark mass at {$T_c^{N_\tau=8}=144$ MeV} in Table \ref{Tab:singcont}.  

\begin{table}
 \centering
 \begin{tabular}{cc}
   \multicolumn{2}{c}{Singular contribution at physical mass}\\[1mm] \hline \rule{0pt}{3ex}
    $\chi_2^B(T_c)$ & $\sim 50\%$ \\[1mm]
    $\chi_2^Q(T_c)$ & $\sim 30\%$ \\[1mm]
    $\chi_2^S(T_c)$ & $\sim 20\%$ \\[1mm] \hline
  \end{tabular}
  \caption{Approximate singular contributions to various $\chi_2^X$ at $T=T_c$ at physical light quark mass $m_l=m_s/27$ at lattice temporal extent $N_\tau=8$.}\label{Tab:singcont}
\end{table}

From the singular contributions estimated above, one can get an idea about the curvature of the chiral transition line in the chiral limit. The ratios of the curvature coefficients $\kappa_2^X$ for different chemical potentials can be obtained directly from the ratio of the corresponding singular contributions as everything else cancels out (see Eq. \ref{Eq:chi2atTc}; note that $f_f^{(1)}(0)$ is a constant universal number). The preliminary estimates of the ratios ${\kappa_2^Q}/{\kappa_2^B}$ and ${\kappa_2^B}/{\kappa_2^S}$ are {2.6 and 1.0} respectively. These values are quite close to the corresponding ratios of the curvatures of the crossover line at physical quark masses, $1.8(8)$ and $0.9(4)$ respectively, obtained in Ref.~\cite{Bazavov:2018mes}. {This indicates that the curvature of the crossover line remains almost unchanged as one goes towards the chiral limit.}

\begin{figure}[t]
\centerline{%
\includegraphics[width=6cm]{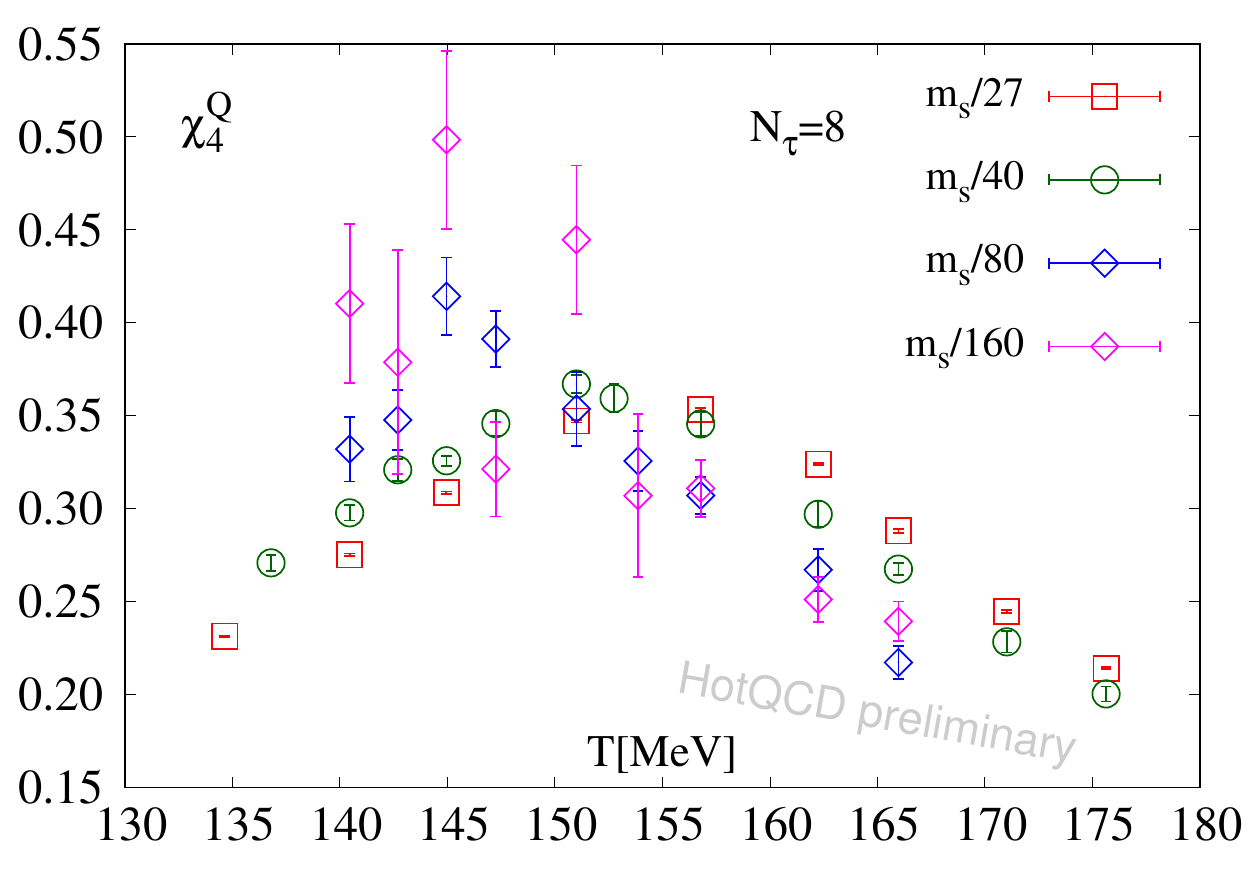}
\includegraphics[width=6cm]{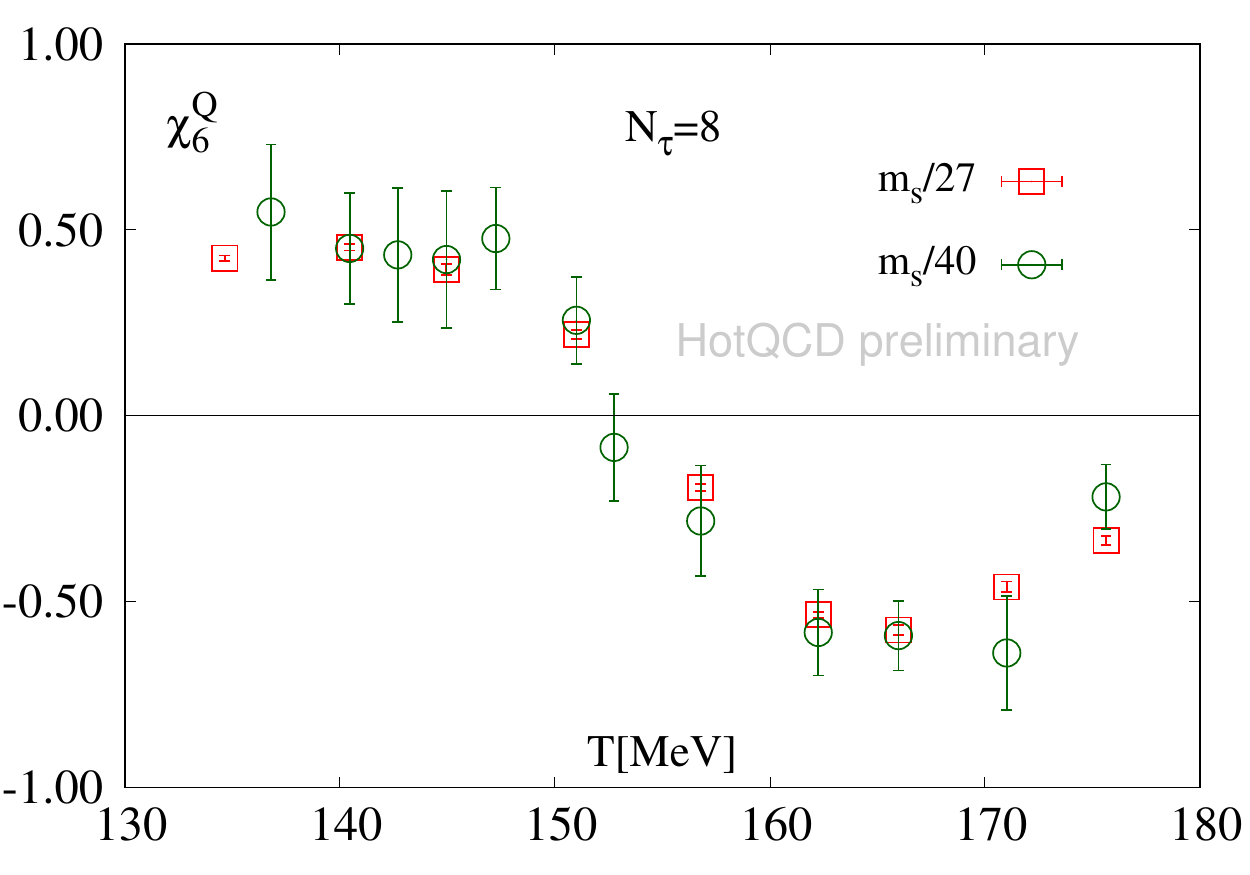}}
\caption{(\textit{Left}) The fourth order cumulant of electric charge fluctuations, $\chi_4^Q$ as a function of temperature for different $H$ values. (\textit{Right}) The sixth order cumulant of electric charge fluctuations, $\chi_6^Q$ versus temperature at two $H$ values.}
\label{Fig:chi4chi6}
\end{figure}

Next, we discuss the qualitative features of the $4^{\text{th}}$ order conserved charge fluctuations. The singular part of $\chi_4^X$ would vanish at $T_c$ as $h^{0.01}$ with $O(2)$ exponents and thus, no divergence would be present in the chiral limit. 
The appearance of the characteristic spike in the full quantity, as seen just for the singular part in the left plot of Fig. \ref{Fig:o4func}, is therefore, dependent on the relative size of the regular term at a given light quark mass $m_l$. Our preliminary results for the fourth order electric charge fluctuation $\chi_4^Q$ as a function of temperature is shown in the left plot of Fig. \ref{Fig:chi4chi6}. 
With decreasing light quark mass $m_l$, a spike seems to be developing near $T_c$. The plot for $\chi_4^B$ shows similar features but is noisy and requires more statistics. The similar plot for strangeness, however, is quite different with a monotonically increasing behavior. This happens most likely due to a relatively large contribution of the regular terms, as seen already in Table \ref{Tab:singcont} for $\chi_2^S$.

According to $O(2)$ or $O(4)$ universality, the $6^{\text{th}}$ and higher order fluctuations are divergent in the chiral limit and should provide a strong evidence for criticality. Due to a diverging singular part, the regular terms should be less relevant towards the chiral limit. 
However, these observables require large statistics and lattice calculations are increasingly expensive at smaller masses. 
Our preliminary result for $\chi_6^Q$ in the right plot of Fig. \ref{Fig:chi4chi6} shows features similar to that of the universal singular term in the right plot of Fig. \ref{Fig:o4func}. 
At first glance, it might look like there is no divergence with decreasing mass in the figure but one must keep in mind that firstly, the regular parts can still be appreciable at these masses and second, the ratio of the peak heights for $H=1/27$ and $H=1/40$ expected from $O(2)$ singular parts is only about $1.26$, which is not too far from the actual data. 
The relatively high positive part at lower $T$ compared to the negative peak at high $T$ for $\chi_6^Q$, in contrast to Fig. \ref{Fig:o4func} (right), is probably due to large regular contributions to the electric charge fluctuations from pions as predicted from hadron resonance gas (HRG) calculations.

\begin{figure}[t]
\centerline{%
\includegraphics[width=6cm]{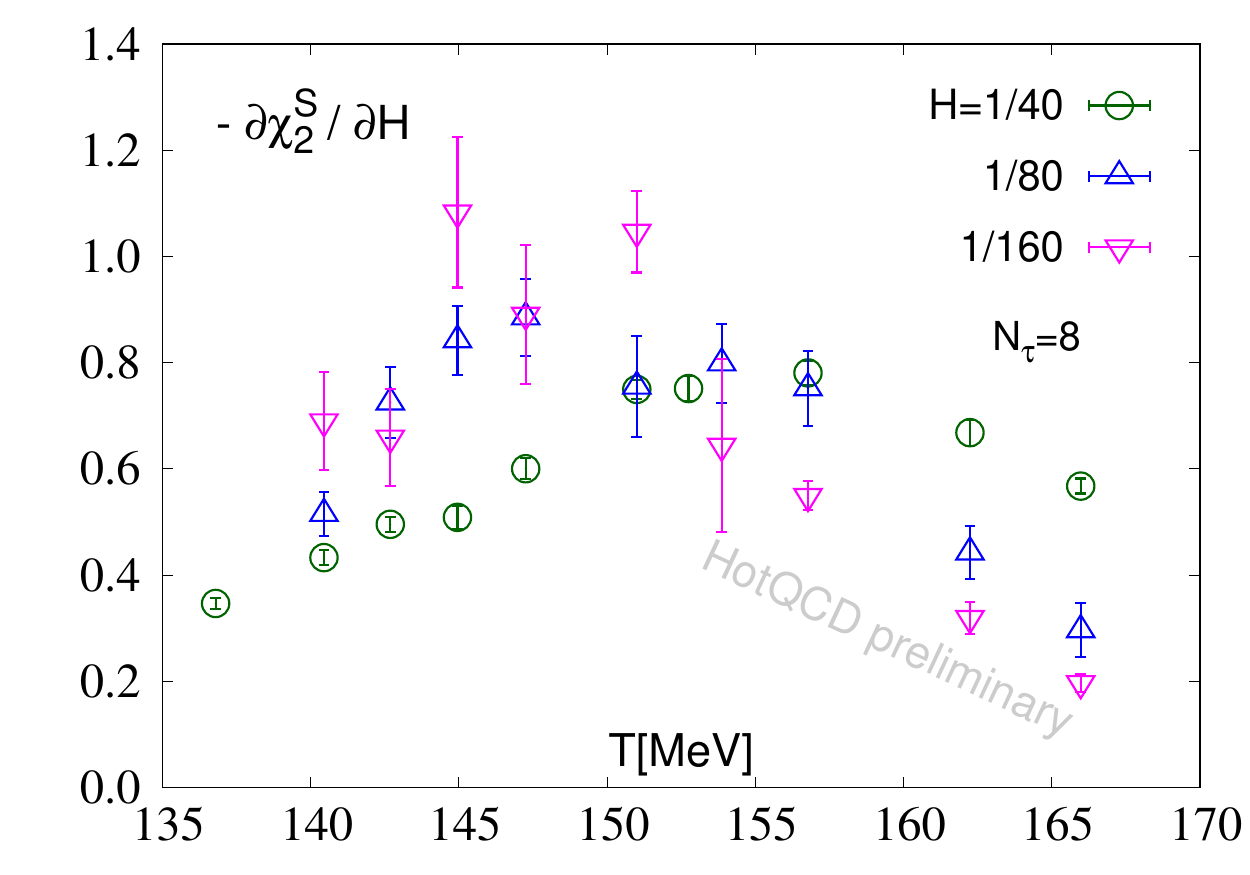}
\includegraphics[width=6cm]{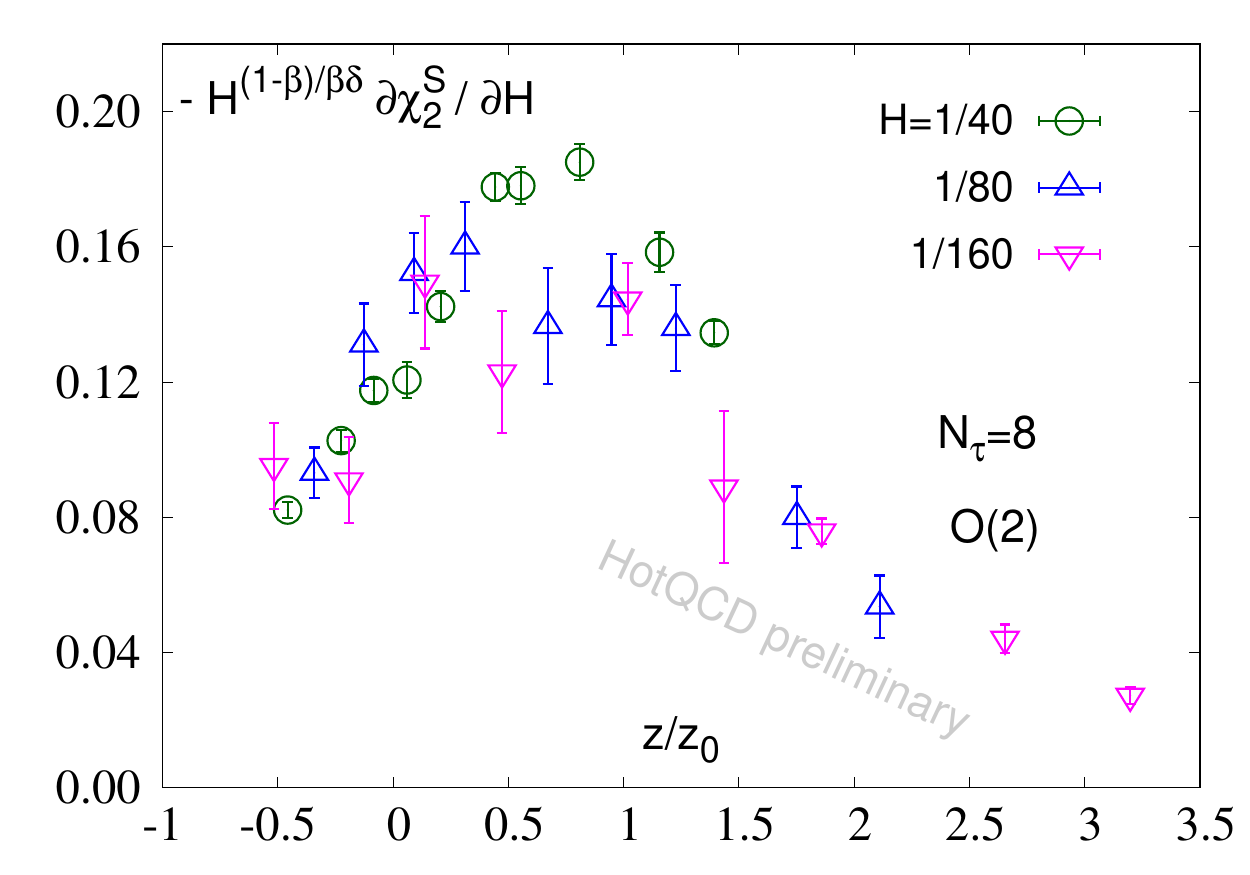}}
\caption{(\textit{Left}) The $H$-derivative of second order strangeness fluctuation $\chi_2^S$ as a function of temperature at different masses. (\textit{Right}) The derivative, rescaled with $H^{(1-\beta)/\beta\delta}$, plotted as a function of $z/z_0 = H^{-1/\beta\delta}(T-T_c)/T_c$ for fixed values of $H$. Apart from regular contributions, the rescaled quantity is proportional to the universal scaling function $f_G^\prime(z)$.}
\label{Fig:chi2s_mder}
\end{figure}
\begin{figure}[t]
\centerline{%
\includegraphics[width=6cm]{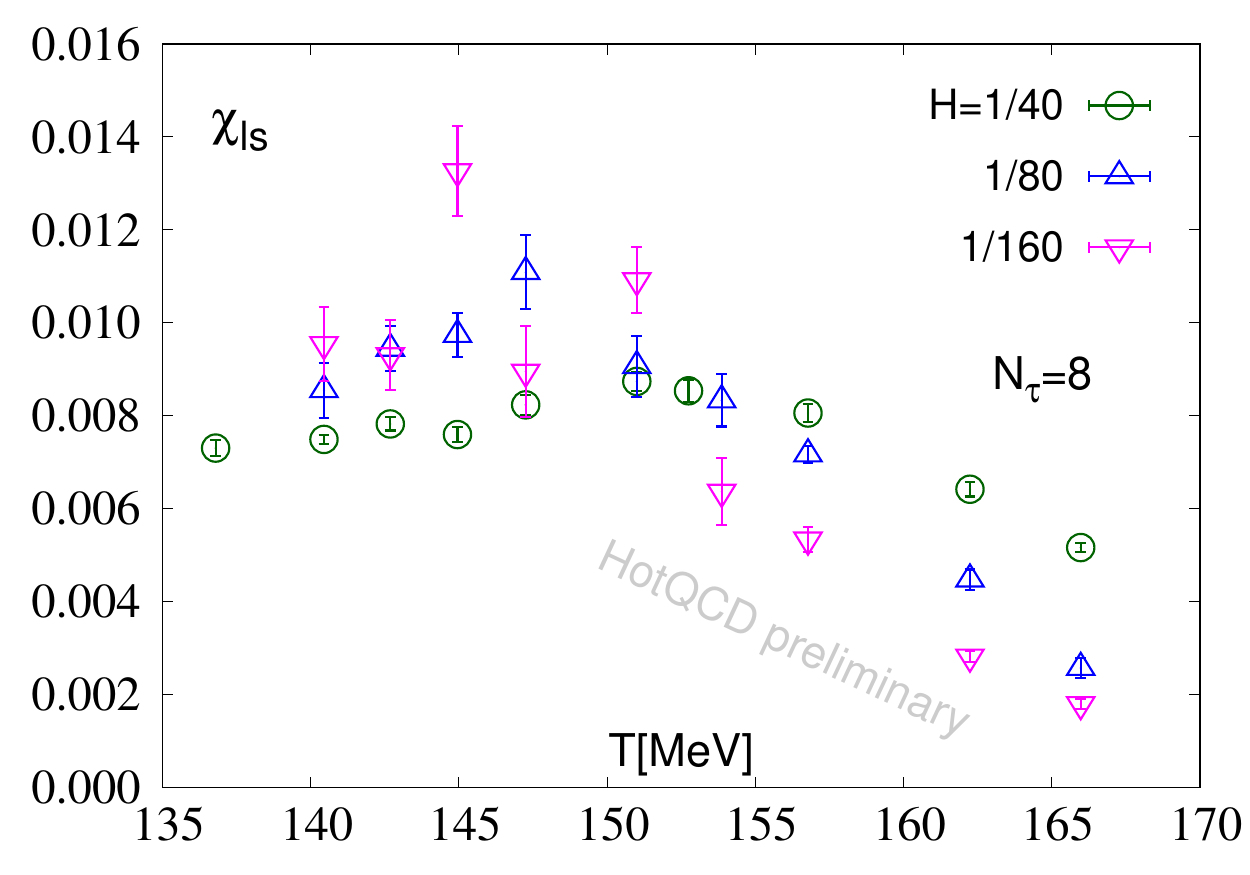}
\includegraphics[width=6cm]{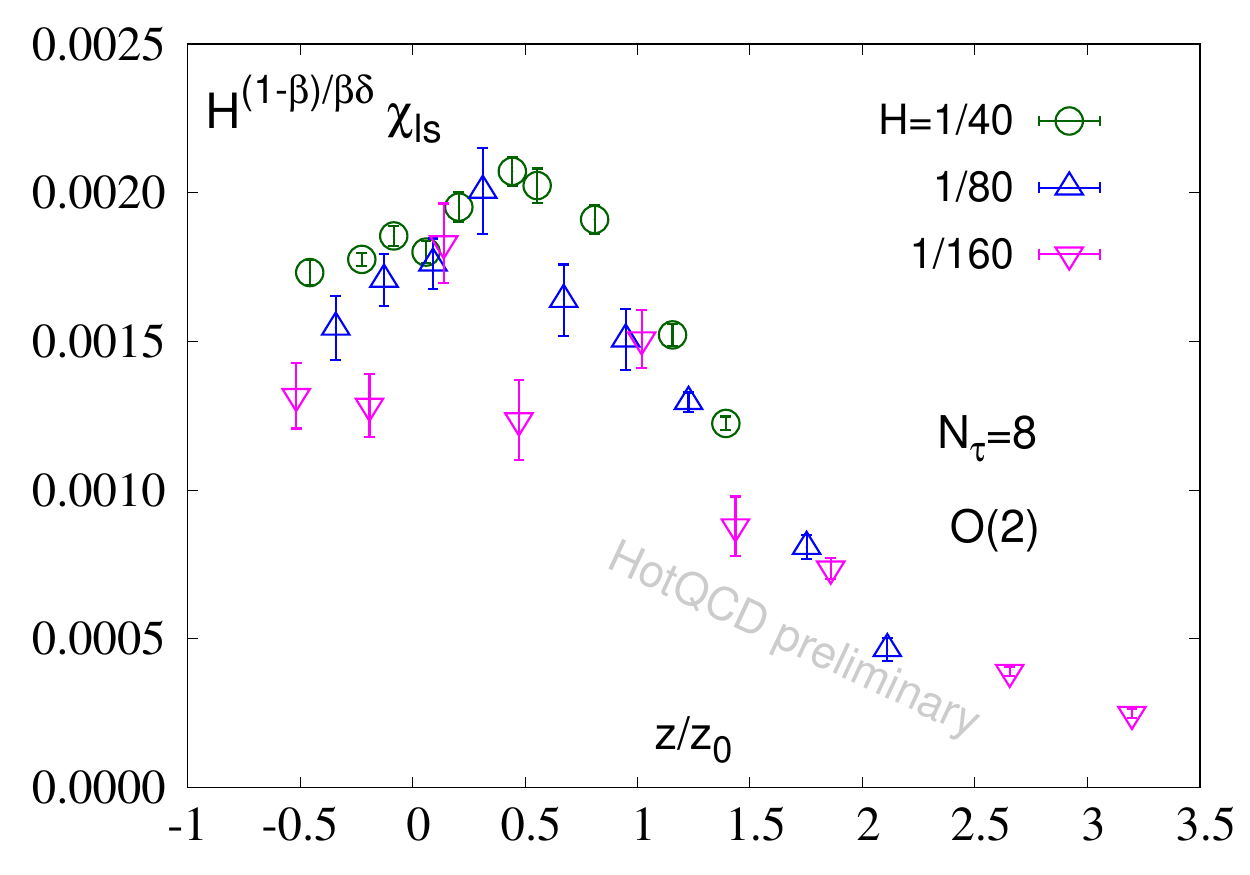}}
\caption{Same as Fig. \ref{Fig:chi2s_mder} but for $H$-derivative of strange chiral condensate, $\chi_{ls}\equiv m_s\partial \expval{\bar{\psi}\psi}_s/\partial m_l$.}
\label{Fig:chi_ls}
\end{figure}
Finally, we discuss results from two mixed observables: $\partial \chi_2^S/\partial H$, the light quark mass derivative of the strangeness fluctuation $\chi_2^S$, and $\chi_{ls}\equiv m_s \partial \expval{\bar{\psi}\psi}_s/\partial m_l$, the mass derivative of the strange quark chiral condensate ({the dimensionless strange chiral condensate is defined as $\expval{\bar{\psi}\psi}_s=\frac{1}{4}\tr M_s^{-1}$, where $M_s$ is the strange quark, staggered fermion matrix}). Both of these quantities have the same divergence in the singular part as discussed towards the end of Sec.~\ref{sec2} but with different non-universal factors. The regular terms in these quantities are also undoubtedly different. It is worthwhile to point out again that the strange quark condensate does not break the 2-flavor chiral symmetry and hence is a energy-like observable. {We show these two susceptibilities in the left plots of Fig. \ref{Fig:chi2s_mder} and \ref{Fig:chi_ls}. An interesting way to look at the scaling behavior is to rescale the quantities with $H^{-(\beta-1)/\beta\delta}$ and plot it against the scaling variable $z$, as shown in Fig.~\ref{Fig:chi2s_mder} (right) and \ref{Fig:chi_ls} (right). Apart from regular contributions, the rescaled observable is proportional to the universal scaling function $f_G^\prime(z)$ (see Eq.~\ref{Eq:mixed}). The data seems to fall on top of each other in the figures with some deviations which can be taken care by considering regular terms. This clearly shows the universal scaling behavior in these quantities already at physical quark masses.} Although seemingly different, quantities like $\chi_{mP}$, the $H$-derivative of the Polyakov loop studied in Ref.~\cite{Clarke:2020htu}, behave exactly same as the mixed susceptibilities discussed above, as expected from universal scaling behavior. It is possible to do a similar analysis of the regular and singular parts of the above-mentioned mixed quantities, as has been done in Ref.~\cite{Clarke:2020htu}.

\section{Conclusions and Outlook} \label{Sec5}
To summarize, the fluctuations of conserved charges at finite lattice spacing seem to be consistent with chiral
phase transition belonging to $O(2)$ universality class, and therefore, to $O(4)$ in the continuum limit. They exhibit
expected energy-like behavior with respect to chiral
phase transition. Even strangeness fluctuations and strange quark condensate 
behave as energy-like quantities in the 2-flavor chiral limit.
The singular contributions can be estimated for different observables 
and may be used to
determine the curvature coefficients of the chiral critical line.
In particular, our analysis for the second order fluctuations at physical pion masses 
show a considerable singular contribution.
We intend to do a future comparison with HRG at smaller masses to understand the 
interplay of singular and regular parts.
A more quantitative understanding of the mixed susceptibilities and conserved charge susceptibilities will be obtained in future through a scaling analysis following Ref.~\cite{Clarke:2020htu}. The work is under progress and 
more statistics are being generated at lower masses in order to achieve proper
continuum and thermodynamic limits.

\section{Acknowledgments}
This work was supported by the Deutsche Forschungsgemeinschaft
(DFG) Proj. No. 315477589-TRR 211; the German Bundesministerium f\"ur Bildung und Forschung through
Grant No. 05P18PBCA1 and the EU H2020-MSCA-ITN-2018-813942 (EuroPLEx). 
We thank the HotQCD Collaboration for providing access to their latest data
sets and for useful discussions.

\bibliographystyle{h-physrev}
\bibliography{bibliography}

\end{document}